\documentclass{kapproc} 

\usepackage{procps} 

\usepackage[dvips]{graphicx}

\upperandlowercase

\normallatexbib 

\def\a{\alpha}
\def\b{\beta}

\def\iomn{i\omega_n}
\def\inun{i\nu_n}
\def\vq{{\bf q}}
\def\vk{{\bf k}}
\def\vr{{\bf r}}
\def\vR{{\bf R}}

\def\hro{\hat{\rho}}
\def\bra{\langle}
\def\ket{\rangle}
\def\cG0{{\cal G}_0}
\def\GRRp{G^{\vR\vR'}}
\def\WRRp{W^{\vR\vR'}}
\def\GRR{G^{\vR\vR}}
\def\WRR{W^{\vR\vR}}
\def\cU{{\cal U}}
\def\cG{{\cal G}}
\def\GH{{G_H}}
 
\begin{document}

\articletitle{Electronic Structure of Strongly Correlated Materials:
towards a First Principles Scheme}


\author{Silke Biermann}
\affil{Centre de Physique Theorique, 
\\
Ecole Polytechnique,
91128 Palaiseau, France}
\email{biermann@cpht.polytechnique.fr}

\author{Ferdi Aryasetiawan}
\affil{Research Institute for Computational Sciences, AIST,
\\
1-1-1 Umezono, Tsukuba Central 2, Ibaraki 305-8568, Japan
}
\email{f-aryasetiawan@aist.go.jp}

\author{Antoine Georges}
\affil{Centre de Physique Theorique, 
\\
Ecole Polytechnique,
91128 Palaiseau, France}
\email{georges@cpht.polytechnique.fr}

\begin{abstract}
We review a recent proposal of a first principles
approach to the electronic structure of materials
with strong electronic correlations.
The scheme combines the GW method with dynamical
mean field theory,
which enables one to treat strong interaction effects.
It allows for a parameter-free description of Coulomb
interactions and screening, and thus avoids the 
conceptual problems inherent
to conventional ``LDA+DMFT'', such as Hubbard
interaction {\it parameters} and double counting terms.
We describe the application of a simplified version of the approach
to the electronic structure of nickel yielding encouraging
results.
Finally, open questions and further perspectives
for the development of the scheme are discussed.
\end{abstract}


\section*{Introduction}

The development of density functional theory in combination
with rapidly increasing computing power has led to revolutionary
progress in electronic structure theory over the last 40 years.
Nowadays, calculations for materials with large unit cells
are feasible, and applications to biological systems and
complex materials of high technological importance are within reach.

The situation is less favorable, however, for so-called
strongly correlated materials, where strong localization and
Coulomb interaction effects cause density functional theory
within the local density approximation (DFT-LDA) to fail.
These are typically materials with partially filled d- or
f-shells. Failures of DFT-LDA reach from missing
satellite structures in the spectra (e.g. in transition metals) 
over qualitatively
wrong descriptions of spectral properties (e.g. certain
transition metal oxides) to severe qualitative errors
in the calculated equilibrium lattice structures (e.g.
the absence of the Ce $\alpha-\gamma$ transition in LDA
or the $30\%$ error on the volume of $\delta$-Pu).
While the former two situations may be at least partially
blamed on the use of a ground state theory in the forbidden 
range of excited states properties, in the latter cases
one faces a clear deficiency of the LDA.
We are thus in the puzzling situation of being able to
describe certain very complex materials, heading for a
first principles description of biological systems, while
not having successfully dealt with others that have much
simpler structures but resist an LDA treatment due to
a particular challenging electronic structure.
The fact that many of these strongly correlated materials
present unusual magnetic, optical or transport properties
has given additional motivation to design electronic 
structure methods beyond the LDA.

Both, LDA+U \cite{anisimov-ldau,anisimov0,sasha-ldau,anisimov}
and LDA+DMFT \cite{anisimov1,sasha-dmft,anisimov2,kotliar-lecture}
techniques are based on an
Hamiltonian that explicitly corrects the LDA description
by corrections stemming from local Coulomb
interactions of partially localized electrons. This
Hamiltonian is then solved within a static or a dynamical
mean field approximation in LDA+U or LDA+DMFT respectively.
In a number of magnetically or orbitally ordered insulators the
LDA underestimation of the gap is successfully corrected
by LDA+U 
(e.g. late transition metal oxides and some rare
earth compounds); however its description of low energy
properties is too crude to describe correlated metals,
where the dynamical character of the mean field is crucial
and LDA+DMFT thus more successful.
Common to both approaches is the need to determine the
Coulomb parameters from independent (e.g. ``constrained LDA'') 
calculations or to fit them to experiments.
Neither of them thus describes long-range Coulomb interactions
and the resulting screening from first principles.
This has led to a recent proposal \cite{gwdmft} of a 
first principles electronic structure method, dubbed
``GW+DMFT'' that we review in this article. Similar
advances have been presented in a model context  
\cite{sun}.
 
The paper is organized as follows: in section 1 we 
give a short overview over the parent theories (GW, DMFT, 
and LDA+DMFT), while in section 2 we introduce
a formal way of constructing approximations by means of a
free energy functional. The form of this functional defining
the GW+DMFT scheme is discussed in section 3; section 4 
presents different conceptual as well as technical issues
related to this scheme. Finally we present results of a
preliminary static implementation combining GW and DMFT,
and conclude the paper with some remarks on further perspectives
for the development of the GW+DMFT scheme.

\section{The parent theories}

\subsection{The \emph{GW }Approximation}

Even if density functional theory is strictly only applicable
to ground state properties, band dispersions of sp-electron
semi-conductors and insulators have been found to be surprisingly
reliable -- apart from a systematic underestimation of band
gaps (e.g. by $\sim 30\%$ in Si and Ge).

This underestimation of bandgaps has prompted a number
of attempts at improving the LDA. Notable among these is the GW approximation
(GWA), developed systematically by Hedin in the early sixties \cite{hedin}. He
showed that the self-energy can be formally expanded in powers of the screened
interaction $W$, the lowest term being $iGW,$ where $G$ is the Green function.
Due to computational difficulties, for a long time the applications of the GWA
were restricted to the electron gas. With the rapid progress in computer
power, applications to realistic materials eventually became possible about
two decades ago. Numerous applications to semiconductors and insulators reveal
that in most cases the GWA \cite{ferdi,aulbur,onida} 
removes a large fraction of the
LDA band-gap error. Applications to alkalis show band narrowing from the LDA
values and account for more than half of the LDA error (although controversy
about this issue still remains \cite{ku}).

The GW approximation relies on Hedin's equations \cite{hedin}, which
state for the self-energy that
\begin{equation}
\Sigma(1,2)=-i\int d3\;d4\;v(1,4)G(1,3)\frac{\delta G^{-1}(3,2)}{\delta
\phi(4)} \label{Sigma}%
\end{equation}
where $v$ is the bare Coulomb interaction, $G$ is the Green function and
$\phi$ is an external time-dependent probing field. We have used the
short-hand notation $1=(x_{1}t_{1})$. From the equation of motion of the Green function%

\begin{equation}
G^{-1}=i\frac{\partial}{\partial t}-H_{0}-\Sigma
\end{equation}%

\begin{equation}
H_{0}=h_{0}+\phi+V_{H}%
\end{equation}
$h_{0}$ is the kinetic energy and $V_{H}$ is the Hartree potential. We then
obtain
\begin{eqnarray}
\frac{\delta G^{-1}(3,2)}{\delta\phi(4)}  &  =-\delta(3-2)\left[
\delta(3-4)+\frac{\delta V_{H}(3)}{\delta\phi(4)}\right]  -\frac{\delta
\Sigma(3,2)}{\delta\phi(4)}\nonumber\\
&  =-\delta(3-2)\epsilon^{-1}(3,4)-\frac{\delta\Sigma(3,2)}{\delta\phi(4)}
\label{dG-1}%
\end{eqnarray}
where $\epsilon^{1}$ is the inverse dielectric matrix. The GWA is obtained by
neglecting the vertex correction $\delta\Sigma/\delta\phi$, which is the last
term in (\ref{dG-1}). This is just the random-phase approximation (RPA) for
$\epsilon^{-1}.$ This leads to%

\begin{equation}
\Sigma(1,2)=iG(1,2)W(1,2)
\end{equation}
where we have defined the screened Coulomb interaction $W$ by%

\begin{equation}
W(1,2)=\int d3v(1,3)\epsilon^{-1}(3,2)
\end{equation}
The RPA dielectric function is given by%

\begin{equation}
\epsilon=1-vP
\end{equation}
where%

\begin{eqnarray}
P(\mathbf{r,r}^{\prime};\omega)  &=& -2i\int\frac{d\omega^{\prime}}{2\pi
}G(\mathbf{r,r}^{\prime};\omega+\omega^{\prime})G(\mathbf{r}^{\prime
},\mathbf{r};\omega^{\prime})\nonumber\\
&=& 2\sum_{i}^{occ}\sum_{j}^{unocc}\psi_{i}(\mathbf{r)}\psi_{i}^{\ast
}(\mathbf{r}^{\prime})\psi_{j}^{\ast}(\mathbf{r)}\psi_{j}(\mathbf{r}^{\prime
})\label{PRPA}\\
& & \times\left\{  \frac{1}{\omega-\varepsilon_{j}+\varepsilon_{i}+i\delta
}-\frac{1}{\omega+\varepsilon_{j}-\varepsilon_{i}-i\delta}\right\}
\end{eqnarray}
with the Green function constructed from a one-particle band structure
$\{\psi_{i},\varepsilon_{i}\}$. The factor of 2 arises from the sum over spin
variables. In frequency space, the self-energy in the GWA takes the form%

\begin{equation}
\Sigma(r,r^{\prime};\omega)=\frac{i}{2\pi}\int d\omega^{\prime}e^{i\eta
\omega^{\prime}}G(r,r^{\prime};\omega+\omega^{\prime})W(r,r^{\prime}%
;\omega^{\prime})
\end{equation}
We have so far described the zero temperature formalism. For finite
temperature we have%

\begin{equation}
P(\mathbf{r,r}^{\prime};i\nu_{n})=\frac{2}{\beta}\sum_{\omega_{k}%
}G(\mathbf{r,r}^{\prime};i\nu_{n}+i\omega_{k})G(\mathbf{r}^{\prime}%
,\mathbf{r};i\omega_{k})
\end{equation}%

\begin{equation}
\Sigma(r,r^{\prime};i\omega_{n})=-\frac{1}{\beta}\sum_{\nu_{k}}G(r,r^{\prime
};i\omega_{n}+i\nu_{k})W(r,r^{\prime};i\nu_{k})
\end{equation}

In the Green function language, the Fock exchange operator in the Hartree-Fock
approximation (HFA) can be written as $iGv$. We may therefore regard the GWA
as a generalization of the HFA, where the bare Coulomb interaction $v$ is
replaced by a screened interaction $W$. We may also think of the GWA as a
mapping to a polaron problem where the electrons are coupled to some bosonic
excitations (e.g., plasmons) and the parameters in this model are obtained
from first-principles calculations.

The replacement of $v$ by $W$ is an important step in solids where screening
effects are generally rather large relative to exchange, especially in metals.
For example, in the electron gas, within the GWA exchange and correlation are
approximately equal in magnitude, to a large extent cancelling each other,
modifying the free-electron dispersion slightly. But also in molecules,
accurate calculations of the excitation spectrum cannot neglect the effects of
correlations or screening. The GWA is physically sound because it is
qualitatively correct in some limiting cases \cite{hedin95}.

The success of the GWA in sp materials has prompted further applications to
more strongly correlated systems. For this type of materials the GWA has been
found to be less successful. 
Application to ferromagnetic nickel \cite{ferdini} illustrates some
of the difficulties with the GWA. Starting from the LDA band structure, a
one-iteration GW calculation 
does reproduce the photoemission quasiparticle band structure
rather well, as compared with the LDA one where the 3d band width is too large
by about 1 eV.
However, the too large LDA
exchange splitting (0.6 eV compared with experimental value of 0.3 eV) remains
essentially unchanged. Moreover, the famous 6 eV satellite, which is of course
missing in the LDA, is not reproduced. These problems point to deficiencies in
the GWA in describing short-range correlations since we expect that both
exchange splitting and satellite structure are influenced by on-site
interactions. In the case of exchange splitting, long-range screening also
plays a role in reducing the HF value and the problem with the exchange
splitting indicates a lack of spin-dependent interaction in the GWA: In the
GWA the spin dependence only enters in $G$ but not in $W$.

The GWA rather successfully improves on the LDA errors on the
bandgaps of Si, Ge, GaAs, ZnSe or Diamond, but
for some materials, such as MgO and InN,
a significant error still remains. The reason for the discrepancy
has not been understood well. One possible explanation is that the result of
the one-iteration GW calculation may depend on the starting one-particle band
structure, since the starting Green's function is usually constructed from
the LDA Kohn-Sham orbitals and energies. 
For example, in the case of InN, the starting LDA band structure
has no gap. This may produce a metal-like (over)screened interaction $W$ which
fails to open up a gap or yields a too small gap in the GW calculation. A
similar behaviour is also found in the more extreme case of NiO, where a
one-iteration GW calculation only yields a gap of about 1 eV starting from an
LDA gap of 0.2 eV (the experimental gap is 4 eV) \cite{ferdi-nio,ferdi}.
This problem may be circumvented by performing a partial
self-consistent calculation in which the self-energy from the previous
iteration at a given energy, such as the Fermi energy of the centre of the
band of interest, is used to construct a new set of one-particle orbitals.
This procedure is continued to self-consistency such that the starting
one-particle band structure gives zero self-energy correction
\cite{ferdi-nio,ferdi,faleev}. 
In the case of NiO this procedure improves the band gap considerably to a
self-consistent value of 5.5 eV and at the same time increases the
LDA\ magnetic moment from 0.9 $\mu_{B}$ to about 1.6 $\mu_{B}$ much closer to
the experimental value of 1.8 $\mu_{B}$ . 
A more serious problem, however, is describing
the charge-transfer character of the top of the valence band. 
Charge-transfer insulators
are characterized by the presence of occupied
and unoccupied 3d bands with the oxygen 2p band in between. The gap is then
formed by the oxygen 2p and unoccupied 3d bands, unlike the gap in LDA, which
is formed by the 3d states. A more appropriate
interpretation is to say that the highest valence state is a charge-transfer
state: During photoemission a hole is created in the transition metal site but
due to the strong 3d Coulomb repulsion it is energetically more favourable for
the hole to hop to the oxygen site despite the cost in energy transfer. A
number of experimental data, notably 2p core photoemission resonance, suggest
that the charge-transfer picture is more appropriate to describe the
electronic structure of transition metal oxides.
The GWA, however,
essentially still maintains the 3d character of the top of the valence
band, as in the LDA, and misses the
charge-transfer character dominated by the 2p oxygen hole. 
A more recent calculation using a more refined
procedure of partial self-consistency has also confirmed these results
\cite{faleev}. The problem with the GWA appears to arise from inadequate
account of short-range correlations, probably not properly treated in the
random-phase approximation (RPA).
As in nickel, the
problem with the satellite arises again
in NiO. Depending on the starting band
structure, a satellite may be reproduced albeit at a too high energy. Thus
there is a strong need for improving the short-range correlations in the GWA
which may be achieved by using a suitable approach based on the dynamical
mean-field theory described in the next section.

\subsection{Dynamical Mean Field Theory}\label{dmft}

Dynamical mean field theory (DMFT) \cite{georges} has originally been
developed within the context of \textit{models} for correlated fermions on a
lattice where it has proven very successful for determining the phase diagrams
or for calculations of excited states properties. It is a non-perturbative
method and as such appropriate for systems with any strength of the
interaction. In recent years, combinations of DMFT with band structure theory,
in particular Density functional theory with the local density approximation
(LDA) have emerged \cite{anisimov1,sasha-dmft}.
The idea is to correct for shortcomings of DFT-LDA due to
strong Coulomb interactions and localization (or partial localization)
phenomena that cause effects very different from a homogeneous itinerant
behaviour. Such signatures of correlations are well-known in transition metal
oxides or f-electron systems but are also present in several elemental
transition metals.

Combinations of DFT-LDA and DMFT, so-called ``LDA+DMFT'' techniques
have so far been applied -- with remarkable success --
to transition metals (Fe, Ni, Mn)
and their oxides (e.g. La/YTiO$_{3}$, V$_{2}$O$_{3}$, Sr/CaVO$_{3}$,
Sr$_2$RuO$_4$)
as well as elemental f-electron materials (Pu, Ce)
and their compounds \cite{anisimov2}.
In the most general formulation, one starts from a many-body Hamiltonian of
the form
\begin{eqnarray}
H &=&\sum_{\{im\sigma\}}(H_{im,i^{\prime}m^{\prime}}^{LDA}-H^{dc}%
)a_{im\sigma}^{+}a_{i^{\prime}m^{\prime}\sigma}\label{LDAUham}\\
&+&\frac{1}{2}\sum_{imm^{\prime}\sigma}U_{mm^{\prime}}^{i}n_{im\sigma
}n_{im^{\prime}-\sigma}\nonumber\\
&+&\frac{1}{2}\sum_{im\neq m^{\prime}\sigma}(U_{mm^{\prime}}^{i}%
-J_{mm^{\prime}}^{i})n_{im\sigma}n_{im^{\prime}\sigma},\nonumber
\end{eqnarray}
where $H^{LDA}$ is the effective Kohn-Sham-Hamiltonian derived from a
self-consistent DFT-LDA calculation. This one-particle Hamiltonian is then
corrected by Hubbard terms for direct and exchange interactions for the
``correlated'' orbitals, e.g. $d$ or $f$ orbitals. In order to avoid double
counting of the Coulomb interactions for these orbitals, a correction term
$H^{dc}$ is subtracted from the original LDA Hamiltonian. The resulting
Hamiltonian (\ref{LDAUham}) is then treated within dynamical mean field theory
by assuming that the many-body self-energy associated with the Hubbard
interaction terms can be calculated from a multi-band impurity model.

This general scheme can be simplified in specific cases, e.g. in systems with
a separation of the correlated bands from the ``uncorrelated'' ones, an
effective model of the correlated bands can be constructed; symmetries of the
crystal structure can be used to reduce the number of components of the
self-energy etc.

In this way,
the application of DMFT to \textit{real solids} crucially relies on 
an appropriate definition of the local screened Coulomb interaction
U (and Hund's rule coupling J).
DMFT then assumes that
local quantities such as
for example the local Green's function or self-energy of the solid can be
calculated from a local impurity model, that is one can find a
\textit{dynamical mean field} $\mathcal{G}_{0}$
such that the Green's function calculated from the effective action
\begin{eqnarray}
S  &=& \int_{0}^{\beta}d\tau\sum_{m\sigma}c_{m\sigma}^{\dagger}(\tau
)\mathcal{G}_{0mm^{\prime}\sigma}^{-1}(\tau-\tau^{\prime})c_{m^{\prime}\sigma
}(\tau^{\prime})\nonumber\label{action}\\
&+&\frac{1}{2}\int_{0}^{\beta}d\tau\sum_{mm^{\prime}\sigma}U_{mm^{\prime}%
}n_{m\sigma}(\tau)n_{m^{\prime}-\sigma}(\tau)\nonumber\\
&+&\frac{1}{2}\int_{0}^{\beta}d\tau\sum_{m\neq m^{\prime}\sigma
}(U_{mm^{\prime}}-J_{mm^{\prime}})n_{m\sigma}(\tau)n_{m^{\prime}\sigma}(\tau)
\end{eqnarray}
coincides with the local Green's function of the solid.
This is in analogy to the 
\textit{representability conjecture} of DMFT in the model
context, where one assumes that e.g. the local Green's function
of a Hubbard model can be represented as the Green's function
of an appropriate impurity model with the same U parameter.
In the case of a lattice with infinite coordination number 
it is trivially seen that this conjecture is correct, since
DMFT yields the exact solution in this case.
Also, the model context is simpler from a conceptual point of
view, since the Hubbard U is given from the outset.
For a real solid the situation is somewhat more complicated, 
since in the construction of the
impurity model \textit{long-range} Coulomb interactions are mimicked by
\textit{local} Hubbard parameters.\footnote{For a discussion of the
appropriateness of local Hubbard parameters see \cite{ferdi03}.}
The notion of locality is also needed
for the resolution of the model within DMFT, which approximates the full
self-energy of the model by a local quantity. Applications of DMFT to
electronic structure calculations (e.g. the LDA+DMFT method) are therefore
always defined within a specific basis set using localized basis functions.
Within an LMTO \cite{lmto}
implementation for example locality can naturally be defined as
referring to the same muffin tin sphere. This amounts to defining matrix
elements $G_{L\mathbf{R},L^{\prime}\mathbf{R}^{\prime}}(i\omega)$ of the full
Green's function
\begin{eqnarray}
\label{Gexpan}
G(\mathbf{r},\mathbf{r}^{\prime},i\omega)=\sum_{LL^{\prime}\mathbf{R}%
\mathbf{R}^{\prime}}\chi_{L\mathbf{R}}^{\ast}(\mathbf{r})G_{L\mathbf{R}%
,L^{\prime}\mathbf{R}^{\prime}}(i\omega)\chi_{L^{\prime}\mathbf{R^{\prime}}%
}(\mathbf{r}^{\prime})
\end{eqnarray}
and assuming that its local, that is ``on-sphere'' part equals the Green's
function of the local impurity model (\ref{action}). Here $\mathbf{R}%
,\mathbf{R}^{\prime}$ denote the coordinates of the centres of the muffin tin
spheres, while $\mathbf{r}$, $\mathbf{r}^{\prime}$ can take any values. The
index $L=(n,l,m)$ regroups all radial and angular quantum numbers. The
dynamical mean field $\mathcal{G}_{0}$ in (\ref{action}) has to be determined
in such a way that the Green's function $G_{impurityL,L^{\prime}}$ of
the impurity model Eq.(\ref{action}) coincides with $G_{L\mathbf{R},L^{\prime
}\mathbf{R}}(i\omega)$ if the impurity model self-energy is used as
an estimate for the true self-energy of the solid. This self-consistency
condition reads
\[
G_{impurity}(i\omega_{n})=\sum_{k}\left(  i\omega_{n}+\mu-H_{o}(k)-\Sigma
(i\omega_{n})\right)  ^{-1}%
\]
where $\Sigma,H_{0}$ and $G$ are matrices in orbital and spin space, and
$i\omega+\mu$ is a matrix proportional to the unit matrix in that space.

Together with (\ref{action}) this defines the DMFT equations that have to be
solved self-consistently. Note that the main approximation of DMFT is hidden
in the self-consistency condition where the local self-energy has been
promoted to the full lattice self-energy.

The representability assumption can actually be extended to other quantities
of a solid than its local Green's function and self-energy. In ``extended
DMFT'' \cite{si,kotliar,kajueter,sengupta} e.g. a two particle correlation
function is calculated and can then be used in order to represent the local
screened Coulomb interaction $W$ of the solid. This is the starting point of
the ``GW+DMFT'' scheme described in section 6.

Despite the huge progress made in the understanding of the electronic
structure of correlated materials thanks to such LDA+DMFT schemes, certain
conceptual problems remain open: These are related to the choice of the
Hubbard interaction parameters and to the double counting corrections. An a
priori choice of which orbitals are treated as correlated and which orbitals
are left uncorrelated has to be made, and the values of U and J have to be
fixed. Attempts of calculating these parameters from constrained LDA
techniques are appealing in the sense that one can avoid introducing external
parameters to the theory, but suffer from the conceptual drawback in that
screening is taken into account in a static manner only \cite{ferdi03}.
Finally, the double counting terms are necessarily ill defined due to the
impossibility to single out in the LDA treatment contributions to the
interactions stemming from specific orbitals. These drawbacks of LDA+DMFT
provide a strong motivation to attempt the construction of an electronic
structure method for correlated materials beyond combinations of LDA and DMFT.

\section{The $\Psi$-functional}
As noted in \cite{almbladh, chitra},
the free energy of a solid can be viewed as a functional 
$\Gamma[G,W]$ of the
Green's function $G$ and the screened Coulomb interaction $W$.
The functional $\Gamma$ can trivially be split into a Hartree
part $\Gamma_H$ and a many body correction $\Psi$, which
contains all corrections beyond the Hartree approximation~:
$\Gamma =  \Gamma_H + \Psi$. The latter 
is the sum of all skeleton
diagrams that are irreducible with respect to both, one-electron
propagator and interaction lines.
 $\Psi[G,W]$ has the following properties:
\begin{eqnarray}\label{eq:def_sigma_P} 
\frac{\delta \Psi}{\delta G} = \Sigma^{xc}
\nonumber
\\
\frac{\delta \Psi}{\delta W} = P.
\end{eqnarray}
We present in the following a different derivation than the one
given in \cite{almbladh}.

We start from the Hamiltonian describing interacting electrons
in an external (crystal) potential $v_{ext}$~:
\begin{eqnarray}
H = H_0 + \frac{1}{2}\sum_{i \ne j} V_{ee}({\bf r}_i - {\bf r}_j)
\end{eqnarray}
with
\begin{eqnarray}
H_0 = - \frac{1}{2} \sum_i \left(\nabla_{i}^2 + v_{ext} \right).
\end{eqnarray}
The electron-electron interaction  
$V_{ee}({\bf r}_i - {\bf r}_j)$
will later on be assumed to be a Coulomb potential.
With the action
\begin{eqnarray}\label{Seff}
S &=& - \int d\tau \Psi^{\dagger} (\delta_{\tau} - H_0 -
V_{Hartree}) \Psi 
\nonumber
\\
&+& \frac{1}{2}\int d\tau~:n~({\bf r}):V~({\bf r} -
{\bf r}^{\prime}):n~({\bf r}^{\prime}):
\end{eqnarray}
the partition function of this system reads~:
\begin{eqnarray}
Z =  \int \mathbf{D}\Psi \mathbf{D}\Psi^{\dagger}
exp(-S)
\end{eqnarray}
Here the double dots denote normal ordering (i.e. Hartree terms
have been included in the first term in \ref{Seff}).
We now do a Hubbard-Stratonovich transform, decoupling Coulomb
interactions beyond Hartree by a continuous field $\phi$, introduce
a coupling constant $\alpha$ for later purposes ($\alpha=1$ 
corresponds to the original problem) and add source terms
coupling to the density fluctuations $\Psi^{\dagger}\Psi$ and the
density of the Hubbard-Stratonovich field respectively~.
The free energy of the system is now a functional of the 
source fields $\Sigma$ and $P$~:
\begin{eqnarray}
F\left[\Sigma, P \right]
= \ln \int \mathbf{D}\Psi \mathbf{D}\Psi^{\dagger} \mathbf{D}\phi
exp{\left(-S[\Sigma, P ]\right)}
\end{eqnarray}
with
\begin{eqnarray}
S [\Sigma, P ] &=& - \int d\tau \Psi^{\dagger} G_{Hartree} \Psi 
+ \frac{1}{2}\int d\tau~ \phi V^{-1} \phi 
\nonumber
\\
&-& i \alpha \int d\tau
\phi(\Psi^{\dagger}\Psi - n)
+ \int d\tau \Sigma \Psi^{\dagger}\Psi + \frac{1}{2} \int d\tau P \phi
\phi
\end{eqnarray}
If in analogy to the usual fermionic Green's function
$G=- \langle \mathbf{T} \Psi \Psi^{\dagger}\rangle$
we define the propagator
$W = \langle \mathbf{T} 
\phi \phi \rangle$
of the Hubbard-Stratonovich field $\phi$, our specific choice
of the coupling of the sources $\Sigma$ and $P$ leads to
\begin{eqnarray}
\frac{\delta F}{\delta \Sigma} = - G 
\end{eqnarray}
and
\begin{eqnarray}
\frac{\delta F}{\delta P} = \frac{1}{2} W
\end{eqnarray}
Performing Legendre transformations with respect to
$G$ and $W/2$ we obtain the free energy as a functional 
of both, the fermionic and bosonic propagators $G$ and $W$
\begin{eqnarray}
\Gamma[G,W] = F + tr(\Sigma G) - \frac{1}{2} tr(PW)
\end{eqnarray}
We note that $W$ can be related to the charge-charge
response function~$\chi(\vr,\vr';\tau-\tau')\equiv \bra
T_\tau[\hro(\vr,\tau)-n(\vr)][\hro(\vr',\tau')-n(\vr')]\ket$~:
\begin{eqnarray}
W(\vr,\vr';\iomn) &=& V (\vr-\vr') 
\nonumber
\\
&-& \int d\vr_1 d\vr_2 V(\vr-\vr_1)\chi(\vr_1,\vr_2;\iomn)V(\vr_2-\vr')
\end{eqnarray}
This property follows directly from the above functional
integral representation and justifies the identification of
$W$ with the screened Coulomb interaction in the solid.
Taking advantage of the coupling constant $\alpha$ introduced
above we find that $\Gamma$ is naturally split into two parts
\begin{eqnarray}\label{gamma}
\Gamma_{\alpha=1}[G,W]= \Gamma_{\alpha=0} + \Psi[G,W]
\end{eqnarray}
where the first is just the Hartree free energy
\begin{eqnarray}
\Gamma_{\alpha=0}(G,W)&=&Tr\ln G-Tr[(G_{H}^{-1}-G^{-1})G]
\nonumber
\\
&-&\frac{1}{2}Tr\ln W+\frac{1}%
{2}Tr[(v^{-1}-W^{-1})W]
\end{eqnarray}
(with 
$G_{H}^{-1}=i\omega_{n}+\mu+\nabla^{2}/2-V_{H}$ 
denoting the Hartree Green's function and $V_{H}$ the
Hartree potential), 
while the
second one
\begin{eqnarray}
\Psi = \int_0^1 d \alpha \frac{\delta \Gamma}{\delta \alpha}
\end{eqnarray}
contains all non-trivial many-body effects.
Needless to mention, this correlation functional $\Psi$ cannot
be calculated exactly, and approximations have to be found.

The GW approximation consists in retaining the first order
term in $\alpha$ only, thus 
approximating the $\Psi$-functional
by
\begin{eqnarray}
\Psi[G,W] = -\frac{1}{2} Tr(GWG).
\end{eqnarray}
We then find trivially
\begin{eqnarray}
\Sigma = \frac{\delta \Psi}{\delta G} = -G W
\end{eqnarray}
\begin{eqnarray}
P = \frac{\delta \Psi}{\delta W} = G G.
\end{eqnarray}
 
\section{GW+DMFT}
Inspired by the description of screening within the
GW approximation and the great successes of DMFT in the
description of strongly correlated materials, the
``GW+DMFT'' method \cite{gwdmft} is constructed
to retain the advantages of both, GW and DMFT, without
the problems associated to them separately.
In the GW+DMFT scheme, the $\Psi$-functional is constructed
from two elements: its local part is supposed to be
calculated from an impurity model as in DMFT, while
its non-local part is given by the non-local part of
the GW $\Psi$-functional:
\begin{eqnarray}\label{Psi}
\Psi\,=\,\Psi_{GW}^{\rm{non-loc}}[\GRRp,\WRRp]
+\Psi_{imp}[\GRR,\WRR]
\end{eqnarray}
Since the strong correlations present in materials with
partially localized d- or f-electrons are expected to
be much stronger in their local components than in their
non-local ones  
the non-local physics is assumed to be well described
by a perturbative treatment as in GW, while local physics
is described within an impurity model in the DMFT sense.
 
More explicitly, the non-local part of the GW+DMFT
$\Psi$-functional is given by
\begin{eqnarray}
\Psi_{GW}^{\rm{non-loc}}[\GRRp,\WRRp]
= \Psi_{GW}[\GRRp,\WRRp] - \Psi_{GW}^{\rm{loc}}[\GRRp,\WRRp]
\end{eqnarray}
while the local part is taken to be an impurity model 
$\Psi$ functional.
Following (extended) DMFT, this onsite part of the functional is generated
from a local {\it quantum impurity problem} (defined on a single
atomic site).
The expression for its free energy functional
$\Gamma_{imp}[G_{imp},W_{imp}]$
is analogous to (\ref{gamma}) with 
${\cal G}$ replacing $\GH$ and ${\cal U}$ replacing $V$~:
\begin{eqnarray}
\Gamma_{imp}[G_{imp},W_{imp}]
&=&Tr\ln G_{imp}-Tr[({\cal G}^{-1}-G_{imp}^{-1})G_{imp}]
\nonumber
\\
&-&\frac{1}{2}Tr\ln W_{imp}+\frac{1}%
{2}Tr[({\cal U}^{-1}-W_{imp}^{-1})W_{imp}]
\nonumber
\\
&+&\Psi_{imp}\lbrack G_{imp},W_{imp} \rbrack \label{LWimp}%
\end{eqnarray}
The impurity quantities $G_{imp},W_{imp}$ can thus be
calculated from the
effective action:
\begin{eqnarray}\label{eq:action}\label{Simp}
&S&=\int d\tau 
d\tau' \left[ -
\sum 
c^{\dagger}_{L}(\tau) {\cal G}^{-1}_{LL'}(\tau-\tau')c_{L'}(\tau')
\right.
\\ \nonumber
&+&\frac{1}{2}
\left.
\sum 
:c^\dagger_{L_1}(\tau)c_{L_2}(\tau):\cU_{L_1L_2L_3L_4}(\tau-\tau')
:c^\dagger_{L_3}(\tau')c_{L_4}(\tau'): \right]
\hskip-1cm
\end{eqnarray}
where the sums run over all orbital indices $L$.
In this expression, $c_L^{\dagger}$ is a creation operator associated with orbital $L$
on a given sphere, and the double dots denote normal ordering (taking care of
Hartree terms).

The construction (\ref{Psi})
of the $\Psi$-functional is the only ad hoc
assumption in the GW+DMFT approach. The explicit form of the
GW+DMFT equations follows then directly from the functional relations
between the free energy, the Green's function, the screened
Coulomb interaction etc.
 Taking derivatives of (\ref{Psi}) as in (\ref{eq:def_sigma_P}) it is seen
that the complete self-energy and polarization operators read:
\begin{eqnarray}\label{Sig_a}
\Sigma^{xc}(\vk,\iomn)_{LL'} &=& \Sigma_{GW}^{xc}(\vk,\iomn)_{LL'}
\\
&-& \sum_\vk \Sigma_{GW}^{xc}(\vk,\iomn)_{LL'}
+ [\Sigma^{xc}_{imp}(\iomn)]_{LL'}
\nonumber
\\
\label{P_a}
P(\vq,\inun)_{\a\b} &=& P^{GW}(\vq,\inun)_{\a\b}
\\
&-& \sum_\vq P^{GW}(\vq,\inun)_{\a\b}+ P^{imp}(\inun)_{\a\b}
\nonumber
\end{eqnarray}

The meaning of (\ref{Sig_a}) is transparent:
the off-site part of the self-energy is taken from
the GW approximation, whereas the onsite part is calculated
to all orders
from the dynamical impurity model.
This treatment thus goes beyond usual E-DMFT, where
the lattice self-energy and polarization are just taken
to be their impurity counterparts.
The second term in
(\ref{Sig_a}) substracts the onsite component of the GW
self-energy thus avoiding double counting.
At self-consistency this term can be rewritten as:
\begin{equation}\label{Sig_correction}
\sum_\vk \Sigma_{GW}^{xc}(\tau)_{LL'}= - \sum_{L_1L_1'}
W^{imp}_{LL_1L'L'_1}(\tau) G_{L'_1L_1}(\tau)
\end{equation}
so that
it precisely substracts the contribution of the GW diagram to
the impurity self-energy. Similar considerations apply
to the polarization operator.
 
From a technical point of view, we note that while one-particle
quantities such as the self-energy or Green's function are
represented in the localized basis labeled by $L$, two-particle
quantities such as $P$ or $W$ are expanded in a two-particle
basis, labeled by Greek indices in (\ref{P_a}).
We will discuss this point more in detail below.

We now outline the iterative loop which determines $\cG$ and $\cU$ self-consistently
(and, eventually, the full self-energy and polarization operator):
\begin{itemize}
\item The impurity problem (\ref{Simp}) is solved, for a given choice of $\cG_{LL'}$ and
$\cU_{\a\b}$: the ``impurity'' Green's function
\begin{equation}
G_{imp}^{LL'}\equiv - \langle T_\tau c_L(\tau)c^+_{L'}(\tau')\rangle_S
\end{equation}
is calculated, together with the impurity self-energy
\begin{equation}
\Sigma^{xc}_{imp}\equiv\delta\Psi_{imp}/\delta G_{imp}=\cG^{-1}-G_{imp}^{-1}.
\end{equation}
The two-particle correlation function
\begin{equation}
\chi_{L_1L_2L_3L_4}=\langle :c^\dagger_{L_1}(\tau)c_{L_2}(\tau):
:c^\dagger_{L_3}(\tau')c_{L_4}(\tau'):\rangle_S 
\end{equation}
must also be evaluated.
\item
The impurity effective interaction is constructed as follows:
\begin{equation}\label{eff_inter}
W_{imp}^{\a\b} = \cU_{\a\b} -
\sum_{L_1\cdots L_4}\sum_{\gamma\delta} \cU_{\a\gamma}
O^{\gamma}_{L_1L_2}
\chi_{L_1L_2L_3L_4} [O^{\delta}_{L_3L_4}]^* \cU_{\delta\b}
\end{equation}
where 
$O_{L_1L_2}^\a\equiv\langle\phi_{L_1}\phi_{L_2}|B^\a\rangle$
is the overlap matrix between two-particle states and products
of one-particle basis functions.
The polarization operator of the impurity problem is then obtained as:
\begin{equation} 
P_{imp}\equiv -2\delta\Psi_{imp}/\delta W_{imp} = \cU^{-1}-W_{imp}^{-1},
\end{equation} 
where all  
matrix inversions are performed in the two-particle basis
$B^\a$
(see the discussion at the end of this section).
\item
From Eqs.~(\ref{Sig_a}) and (\ref{P_a})
the full $\vk$-dependent Green's function $G(\vk,\iomn)$ and
effective interaction $W(\vq,\inun)$ can be constructed. The self-consistency condition
is obtained, as in the usual DMFT context, by requiring that the onsite
components of these quantities coincide with $G_{imp}$ and $W_{imp}$. In practice, this
is done by computing the onsite quantities
\begin{eqnarray}\label{Glocal}
G_{loc}(\iomn) &=& \sum_\vk [\GH^{-1}(\vk,\iomn) - \Sigma^{xc}(\vk,\iomn) ]^{-1}\\
\label{Wlocal}
W_{loc}(\inun) &=& \sum_\vq [V_{\vq}^{-1} - P(\vq,\inun)]^{-1}
\end{eqnarray}
and using them to update the
Weiss dynamical mean field ${\cal G}$
and the impurity model interaction ${\cal U}$ according to:
\begin{eqnarray}\label{update}
\cG^{-1} = G_{loc}^{-1} + \Sigma_{imp}
\\
\label{updateU}
\cU^{-1} = W_{loc}^{-1} + P_{imp}
\end{eqnarray}
\end{itemize}

\begin{figure}
\centerline{\includegraphics[width=12cm,angle=270]{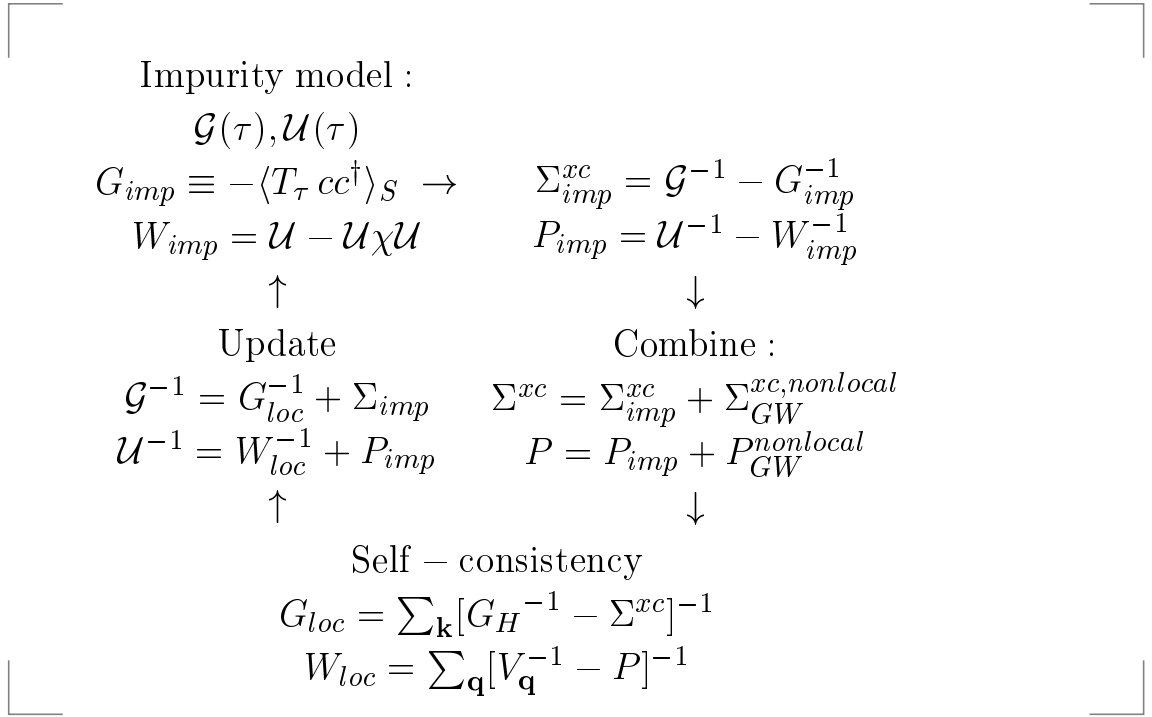}}
\caption{\label{cycle}
Schematic representation of the inner (DMFT) self-consistency
cycle of the GW+DMFT scheme, consisting of the construction
of the impurity model effective action (\ref{Simp}) from
the Weiss field $\cG$ and the impurity Coulomb interaction $\cU$, the solution
of the impurity model leading to an estimate for the impurity
self-energy $\Sigma_{imp}$ and the polarization $P_{imp}$ and 
the self-consistency condition
where local quantities of the solid are calculated and then
used to update the impurity model. Full self-consistency
of the whole scheme requires an additional outer cycle updating
the GW Hartree Potential corresponding to 
the obtained electronic density (cf text).
}
\end{figure}
This cycle (which is summarized in Fig.\ref{cycle})
is iterated until self-consistency for
$\cG$ and $\cU$ is obtained (as well as on $G$, $W$, $\Sigma^{xc}$ and $P$).
Eventually, self-consistency over the local electronic density can also
be implemented,
(in a similar way as in LDA+DMFT \cite{savrasov_pu,savrasov_functional})
by recalculating $\rho(\vec{r})$ from the Green's function at
the end of the convergence cycle above, and constructing an updated Hartree potential.
This new density is used as an input of a new GW calculation,
and convergence over this external loop must be reached.

We stress that in this scheme the Hubbard interaction $\cU$ is
no longer an external parameter but is determined self-consistently.
The appearance of the two functions $\cU$ and $W_{loc}$ might appear 
puzzling at first sight, but has a clear physical interpretation:
$W_{loc}$ is the fully screened Coulomb interaction, while $\cU$ is
the Coulomb interaction screened by all electronic degrees of freedom
that are not explicitly included in the effective action. So for
example onsite screening is included in $W_{loc}$ but not in $\cU$.
From Eq. (\ref{updateU}) it is seen that further screening of $\cU$
by the onsite polarization precisely leads to $W_{loc}$.

In the following we discuss some important issues for the
implementation of the proposed scheme, related to the choice of
the basis functions for one-particle and two-particle quantities.
In practice the self-energy is expanded in some basis set $\{\phi_{L}\}$
localized in a site. The polarization function on the other hand is expanded
in a set of two-particle basis functions $\{\phi_{L}\phi_{L^{\prime}}\}$
(product basis) since the polarization corresponds to a two-particle
propagator. For example, when using the linear muffin-tin orbital (LMTO)
band-structure method, the product basis consists of products of LMTO's. These
product functions are generally linearly dependent and a new set of optimized
product basis (OPB) \cite{ferdi} is constructed by forming linear combinations
of product functions, eliminating the linear dependencies. We denote the OPB
set by $B_{\alpha}=\sum_{LL^{\prime}}\phi_{L}\phi_{L^{\prime}}c_{LL^{\prime}%
}^{\alpha}$. To summarize, one-particle quantities like $G$ and $\Sigma$ are
expanded in $\{\phi_{L}\}$ whereas two-particle quantities such as $P$ and $W$
are expanded in the OPB set $\{B_{\alpha}\}$. It is important to note that the
number of $\{B_{\alpha}\}$ is generally smaller than the number of $\{\phi
_{L}\phi_{L^{\prime}}\}$ so that quantities expressed in $\{B_{\alpha}\}$ can
be expressed in $\{\phi_{L}\phi_{L^{\prime}}\}$, but not vice versa.
E.g. matrix elements in
products of LMTOs can be obtained from those in the $\{B_{\alpha}\}$
basis via the transformation
\begin{equation}\label{trafo}
W_{L_1L_2L_3L_4}^{\vR\vR'}\equiv
\langle\phi_{L_1}^{\vR}\phi_{L_2}^{\vR}|W|\phi_{L_3}^{\vR'}\phi_{L_4}^{\vR'}\rangle =
\sum_{\a\b}O_{L_1L_2}^\a W_{\a\b}^{\vR\vR'} O_{L_3L_4}^{\b *}
\end{equation}
with the overlap
matrix $O_{L_1L_2}^\a\equiv\langle\phi_{L_1}\phi_{L_2}|B^\a\rangle$,
but the knowledge of the matrix elements $W_{L_1L_2L_3L_4}$ alone does not
allow to go back to the $W_{\a\b}$.

\section{Challenges and open questions}
\subsection{Global self-consistency}

As has been pointed out, the above GW+DMFT scheme involves self-con\-sistency
requirements at two levels: for a given density, that is Hartree potential,
the dynamical mean field loop detailed in section \ref{dmft} must be
iterated until $\cG$ and $\cU$ are self-consistently determined.
This results in a solution for $G$, $W$, $\Sigma^{xc}$ and $P$ 
corresponding to this given Hartree potential. Then, a new estimate
for the density must be calculated from $G$, leading to a new estimate
for the Hartree potential, which is reinserted into (\ref{Glocal}).
This external loop is iterated until self-consistency over the local
electronic density is reached.

The external loop is analogous to the one performed when GW calculations
are done self-consistently. From calculations on the homogeneous electron
gas \cite{holm} it is known, that self-consistency at this level 
worsens the description of spectra (and in particular washes out
satellite structures) when vertex corrections are neglected.
Since the combined GW+DMFT scheme, however, includes the local
vertex to all orders and only neglects non-local vertex corrections 
we expect the situation to be more favorable
in this case. Test calculations to validate this hypothesis are an
important subject for future studies. First steps in this
direction have been undertaken in \cite{sun2}.

\subsection{The notion of locality~: the choice of the basis set}

The construction of the GW+DMFT functional $\Psi_{GW}$
crucially relies on the notion of local and nonlocal 
contributions. In a solid these notions can only be
defined by introducing a basis set of localized functions
centered on the atomic positions. Local components are then 
defined to be those functions the arguments of which refer
to the same atomic lattice site. We stress that in this
way the concept of locality is {\it not} a
pointwise ($\delta$-like) one.
It merely means that local quantities have an expansion 
(\ref{Gexpan}) within their atomic sphere 
the form of which is determined by the basis functions
used.
This feature is shared between LDA+DMFT and the GW+DMFT
scheme, and approaches that could directly work in the
continuum are so far not in sight. If however, the
basis set dependence induced by this concept is of
practical importance remains to be tested.

Within LDA+DMFT the choice of the
basis functions enters at two stages~: First,
the definition of the onsite U and the quality of the
approximation consisting in the neglect of offsite
interactions depends on the degree of localization
of the basis functions. Second, 
the DMFT approximation promoting a local quantity
to the full self-energy of the solid is the better
justified the more localized the chosen basis functions
are.
However, in the spirit of obtaining an accurate low-energy
description one might -- depending on the material under
consideration -- in some cases be led to work in a Wannier function
basis incorporating weak hybridisation effects of more
extended states with the localized ones. This then leads
to slightly more extended basis functions, and one has
to compromise between maximally localized orbitals and
an efficient description of low-energy bands.

In GW+DMFT, some non-local corrections to the self-energy
are included, so that the DMFT approximation should
be less severe. More importantly, this scheme
could (via the
$\cU$ self-consistency requirement) automatically adapt
to more or less localized basis functions by choosing
itself the appropriate $\cU$. 
Therefore, the basis set dependence is likely to
be much weaker in GW+DMFT than in LDA+DMFT.

\subsection{Separation of correlated and uncorrelated orbitals}

As mentioned in section 1 the construction of the
LDA+DMFT Hamiltonian requires an {\it a priori} choice of which 
orbitals are treated as correlated or uncorrelated orbitals.
Since in GW+DMFT the Hubbard interactions are determined
self-consistently it might be perceivable not to perform
such a separation at the outset, but to rely on the self-consistency
cycle to find small values for the interaction between itinerant
(e.g. s or p) orbitals. Even if this issue would probably 
not play a role for practical calculations, it would be
satisfactory from a conceptual point of view to be able
to treat all orbitals on an equal footing. It is therefore
also an important question for future studies.

\subsection{Dynamical impurity models}
Dynamical impurity models are hard to solve, since standard techniques
used for the solution of static impurity models within usual DMFT,
such as the Hirsch-Fye QMC algorithm or approximate techniques such
as the iterative perturbation theory are not applicable.
First attempts have been made in \cite{sun} and
\cite{rubtsov} using different auxiliary
field QMC schemes and in \cite{florens} using an
approximate ``slave-rotor'' scheme.
These techniques, however, have so far only been applied in the
context of model systems, and their implementation in a multi-band realistic
calculation is at present still a challenging project.
In the following section we therefore present
a simplified static implementation of the GW+DMFT scheme.

\section{Static implementation}

\begin{figure}
\centerline{\includegraphics[width=8cm]{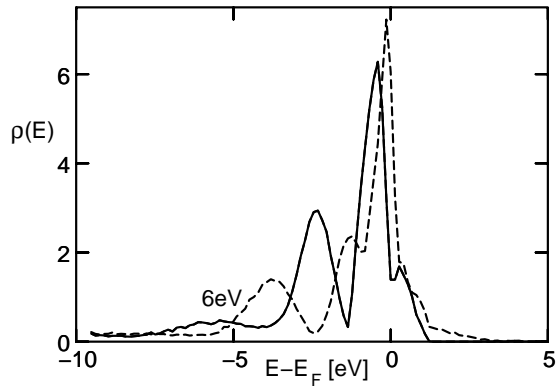}}
\caption{\label{dos}
Partial density of states of d-orbitals of
nickel (solid [dashed] lines give the majority [minority]
spin contribution) as obtained from the
combination of GW and DMFT (see text).
For comparison with LDA and LDA+DMFT results see
\cite{lichtenstein_ni}, 
for experimental spectra see \cite{martensson}.
}
\end{figure}

Here, we demonstrate the feasibility and potential of the approach
within a simplified implementation, which we apply to the electronic
structure of Nickel.
The main simplifications made are:
(i) The DMFT local treatment is applied only to the
$d$-orbitals,
(ii) the GW calculation is done only once, in the form \cite{ferdi}:
$\Sigma^{xc}_{GW} = G_{LDA}\cdot W[G_{LDA}]$, from which the
non-local part of the self-energy is obtained,
(iii) we replace the dynamical impurity problem
by its static limit, solving the impurity model (\ref{Simp})
for a frequency-independent $\cU=\cU(\omega=0)$.
Instead of the Hartree Hamiltonian 
we start from
a one-electron Hamiltonian in the form:
$H_{LDA} - V_{xc,\sigma}^{nonlocal} - \frac{1}{2} \mbox{\bf Tr}
\Sigma_{\sigma}^{imp}(0)$.
The non-local part of this Hamiltonian coincides with that of 
the Hartree Hamiltonian
while its local part is derived from LDA, with a double-counting
correction of the form proposed in
\cite{lichtenstein_ni} in the DMFT context.
With this choice the self-consistency condition
(\ref{Glocal}) reads:
\begin{eqnarray}\label{Glocal2}
G_{loc}^{\sigma}(\iomn) &=& \sum_\vk [
\GH^{-1}(\vk,\iomn)
- (\Sigma^{xc}_{GW})_{non-loc}
\\ \nonumber
&-&(\Sigma_{imp,\sigma} - \frac{1}{2} \mbox{\bf Tr}_{\sigma}
\Sigma_{imp,\sigma}(0) + V_{xc}^{loc})\,]^{-1}
\end{eqnarray}
We have performed finite temperature GW and LDA+DMFT
calculations (within the LMTO-ASA\cite{lmto} with
29 $\vk$-points in the irreducible Brillouin zone) 
for ferromagnetic nickel
(lattice constant 6.654 a.u.), using 4s4p3d4f
states, at the Matsubara frequencies
$\iomn$ corresponding to $T=630 K$, just
below the Curie temperature.
The resulting self-energies are inserted into
Eq.~(\ref{Glocal2}), which is then used to calculate
a new Weiss field according to
(\ref{update}).
The Green's function $G^{\sigma}_{loc}(\tau)$ is recalculated
from the impurity effective action by 
QMC and analytically continued using the
Maximum Entropy algorithm.
The resulting spectral function is plotted in Fig.(\ref{dos}).
Comparison with the LDA+DMFT results in \cite{lichtenstein_ni}
shows that the good description of the satellite structure,
exchange splitting and band narrowing is indeed retained
within the (simplified) GW+DMFT scheme.

\begin{figure}
\centerline{\includegraphics[width=8cm]{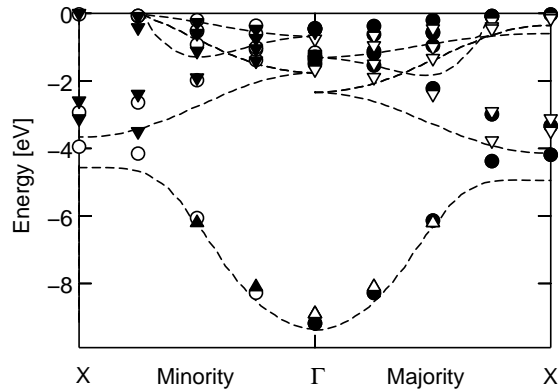}}
\caption{\label{bands}
Band structure of Ni (minority and majority spin) as obtained from
the linearization procedure of the GW+DMFT self-energy
described in the text (dots) in comparison to the
LDA band structure (dashed lines) and the experimental
data of \cite{buenemann} (triangles down) and \cite{martensson}
(triangles up).}
\end{figure}

We have also calculated the quasiparticle band structure,
from the 
poles of (\ref{Glocal2}), after linearization of $\Sigma(\vk,\iomn)$
around the Fermi level
\footnote{Note however that this linearization is no
longer meaningful at energies far away from the Fermi
level. We therefore use the unrenormalized value for
the quasi-particle residue for the s-band ($Z_{s}=1$).}.
Fig.~(\ref{bands}) shows a comparison of GW+DMFT with the
LDA and experimental band structure. It is seen that GW+DMFT correctly
yields the bandwidth reduction compared to the (too large)
LDA value and renormalizes the bands in a
($\vk$-dependent) manner.

We now discuss further the simplifications made in our implementation.
Because of the static approximation (iii), we could not implement self-consistency
on $W_{loc}$ (Eq. (\ref{Wlocal})).
We chose the value of $\cU (\omega=0)$ ($\simeq 3.2 eV$) by calculating
the correlation function $\chi$ and ensuring that
Eq. (\ref{eff_inter}) is fulfilled at $\omega=0$, given the GW value for
$W_{loc}(\omega=0)$ ($\simeq 2.2 eV$ for Nickel \cite{springer}).
This procedure emphasizes the low-frequency, screened value, of the effective
interaction. Obviously, the resulting impurity self-energy
$\Sigma_{imp}$ is then much smaller than the local component of the
GW self-energy (or than $V_{xc}^{loc}$), especially at high frequencies.
It is thus essential to choose the second term in (\ref{Sig_a}) to
be the onsite component of the GW self-energy rather than the
r.h.s of Eq.~(\ref{Sig_correction}). For the same reason, we
included $V_{xc}^{loc}$ in Eq.(\ref{Glocal2}) (or, said differently, we
implemented a mixed scheme which starts from the LDA Hamiltonian for the
local part, and thus still involves a double-counting correction).
We expect that these limitations can be overcome in a self-consistent
implementation with a frequency-dependent $\cU(\omega)$
(hence fulfilling Eq.~(\ref{Sig_correction})). In practice, it might be
sufficient to replace the local part of the GW self-energy
by $\Sigma_{imp}$ for correlated orbitals only. Alternatively, a
downfolding procedure could be used.

\section{Perspectives}

In conclusion, we have reviewed a recent proposal
of an {\it ab initio}
dynamical mean field approach for calculating the
electronic structure of strongly correlated materials,
which combines GW and DMFT.
The scheme aims at avoiding
the conceptual problems inherent to ``LDA+DMFT'' methods, such as
double counting corrections and
the use of Hubbard parameters assigned to correlated orbitals.
A full practical implementation of the GW$+$DMFT scheme is a major goal
for future research, which requires further work on impurity models
with frequency-dependent
interaction parameters \cite{motome_dynamical_QMC,jarrell_holstein,sun}
as well as studies of various possible self-consistency schemes.

\section{Acknowledgments}
 
This work was supported in part by NAREGI Nanoscience Project, Ministry of
Education, Culture, Sports, Science and Technology, Japan and by a grant of
supercomputing time at IDRIS Orsay, France (project number 031393).

\end{document}